\documentclass[usegraphix,usenatbib]{mn2e}

\usepackage{amsmath}
\usepackage{amssymb}
\usepackage{times}
\usepackage{graphicx}
\usepackage{subfigure}
\usepackage{epsfig}
\usepackage{txfonts}

\begin{document}
\voffset-.6in

\pagenumbering{arabic}

\def\rs{{r_{\rm s}}}
\def\thetaa{{\theta_{\rm a}}}
\def\th{{t_{\rm h}}}

\title[Triaxial Cosmological Haloes and the Disc of Satellites]
  {Triaxial Cosmological Haloes and the Disc of Satellites}

\author[Bowden, Evans \& Belokurov]
  {A. Bowden$^1$\thanks{E-mail:adb61, nwe, vasily@ast.cam.ac.uk},
   N.W. Evans$^1$, V. Belokurov$^1$
 \medskip
 \\$^1$Institute of Astronomy, University of Cambridge, Madingley Road,
       Cambridge, CB3 0HA, UK}

\maketitle

\begin{abstract}
We construct simple triaxial generalisations of Navarro-Frenk-White
haloes.  The models have elementary gravitational potentials, together
with a density that is cusped like $1/r$ at small radii and falls off
like $1/r^3$ at large radii. The ellipticity varies with radius in a
manner that can be tailored to the user's specification. The closed
periodic orbits in the planes perpendicular to the short and long axes
of the model are well-described by epicyclic theory, and can be used
as building blocks for long-lived discs.

As an application, we carry out the simulations of thin discs of
satellites in triaxial dark halo potentials. This is motivated by the
recent claims of an extended, thin disc of satellites around the M31
galaxy with a vertical rms scatter of $\sim 12$ kpc and a radial
extent of $\sim 300$ kpc (Ibata et al.  2013). We show that a thin
satellite disc can persist over cosmological times if and only if it
lies in the planes perpendicular to the long or short axis of a
triaxial halo, or in the equatorial or polar planes of a spheroidal
halo. In any other orientation, then the disc thickness doubles on
$\sim 5$ Gyr timescales and so must have been born with an implausibly
small vertical scaleheight.
\end{abstract}


\section{INTRODUCTION}

This paper provides a family of flexible, triaxial, cosmologically
inspired haloes.  There is extensive evidence from numerical
simulations that dark halos are well-represented by
Navarro-Frenk-White (NFW) models with potential-density
pair~\citep[see e.g.,][]{Mo10}
\begin{equation}
\rho_{\rm NFW}(r) = {\rho_0 \rs^3\over r(r+\rs)^2},\qquad \psi_{\rm NFW}(r) =
  4\pi G \rho_0 {\rs^3\over r} \log ( 1+ {r\over\rs}).
\label{eq:nfw}
\end{equation}
Here, $\rs$ and $\rho_0$ are the two parameters of the model providing
a characteristic scalelength and density respectively. The parameters
$\rho_0$ and $\rs$ are known to be related by the mass-concentration
relationship, albeit with some scatter~\citep[see e.g.,][]{Mac07}.

Nonetheless, dark haloes are generically triaxial, whilst the NFW
model is spherical. In purely collisionless simulations, dark haloes
have typical axis ratios $b/a \sim 0.6$ and $c/a \sim 0.4$, where
$c<b<a$ are the short, intermediate and long axes of the triaxial
figure~\citep[see e.g.,][]{Ji02,Al06}. When baryons are included,
haloes remain triaxial, but tend to become rounder and more oblate,
with typical ratios $b/a \sim 0.9$ and $c/a \sim 0.8$ ~\citep[see
  e.g.,][]{Du94,Ka04,De11,Ze12}.  So, it would be useful to have a set
of models based on the Navarro-Frenk-White potential, but which can be
triaxial. The most obvious way of doing this is to take the expression
for $\rho_{\rm NFW}(r)$ and replace $r^2$ by $m^2 = x^2 + y^2/p^2 +
z^2/q^2$, where ($x,y,z$) are Cartesians and $p = b/a$ and $q =c/a $
are the flattenings. This though leads to cumbersome quadratures for
the gravitational potential, and so is unattractive for applications
such as orbit integrations.  The other obvious alternative of
replacing $r^2$ by $m^2$ in the potential $\psi_{\rm NFW}(r)$ leads to
models with negative mass, and so is still less desirable!

Here, we use a classical method that has been known since the
nineteenth century. Spherical harmonics are eigenfunctions of the
angular part of the Laplacian, and so provide a way of building simple
and analytic potential-density pairs.  This technique was popularised
by \citet{Sch79} in his numerical models for triaxial elliptical
galaxies, in which the spherical Hubble profile was modified by the
addition of spherical harmonics. Subsequently, the technique was
exploited by others, primarily with a view to understanding the
intrinsic shapes of elliptical galaxies~\citep[see
  e.g.,][]{He89,Z96,Sch94}. Here, we will exploit the technique to
build triaxial dark haloes.

One reason to develop such models is to assess the longevity of
structures in galaxy haloes. As a particular application, we study the
recent claims of rotationally-supported discs of satellite galaxies in
the M31 halo. This subject goes back at least as far as \citet{Fu95}
who found early hints of discs of satellites. Subsequently,
\citet{Ko06} identified a near-polar great plane containing 9
satellites (M32, NGC 147, PegDIG, And I, And III, And V, And VI, And
VII and And IX). They conjectured that this might be caused either by
tidal break-up of a larger galaxy, or by preferential accretion along
filaments or a prolate dark halo shape. The quality of the data has
been substantially improved with the recent work of
\citet{Co12,Co13}. First, the detection efficiency of the survey was
computed, allaying fears that selection biases may be contributing to
the results. Second, the distances of the now more numerous satellites
were derived by a Bayesian method applied to the Tip of the Red Giant
Branch (TRGB).  \citet{Co13} confirm the existence of a polar plane,
now containing 15 satellites (Ands I, III, IX, XI, XII, XIII, XIV,
XVI, XVII, XXV, XXVI, XXVII, NGC 147, NGC 185 and And XXX). Although
the pole of the plane is $45^\circ$ away from that found by
\citet{Ko06}, the commonality of some of the component satellites
suggests that it may be the same structure.  The plane is aligned with
the Milky Way, and is orthogonal to the plane of the Milky Way disc.
The root mean square (rms) scatter in distance from the plane is $\sim
12.6$ kpc.  Subsequently, \citet{Ib13} analyzed the radial velocities
of the satellites and came to a very remarkable conclusion. They
argued that 13 of the 15 satellites possess a coherent, rotational
motion. This, they suggested, is evidence of a gigantic structure, a
vast rotating plane of satellites, nearly 400 kpc in extent, but very
thin, with an rms scatter of 14.1 kpc at 99 \% confidence level.

The claims of such a thin but extended disc of satellites around M31
are very surprising.  Although the orbits of the satellites have
periods of several Gyr, they are subject to torques from the
quadrupole moment from the M31 bulge, disc and the triaxial dark halo.
They also are subject to perturbing tidal forces from the nearby Milky
Way galaxy. The orbital evolution driven by the combined effects of
these forces might be expected to thicken any thin disc of satellites
quite quickly.  Of course, triaxial potentials can support closed
periodic orbits in the planes perpendicular to the long and the short
axis only.  These are natural orbits on which cold gas or streams of
satellite might settle and persist to give long-lived disc-like
structures. However, even if located in such a propitious plane, it is
far from clear that a thin disc of satellites could persist for long
times.

The paper is arranged as follows. In Section 2, we develop triaxial
halos models generalising the NFW potential. The closed orbits in the
principal planes are found by epicyclic theory. Their properties are
interesting as the orbits can potentially form the basis of long-lived
discs. Section 3 describes our algorithm for seeding the initial
conditions of orbits in extended discs in the outer parts of such
halos, and analyses the results of such simulations. We show
that, in general, a thin disc of satellites can be expected to double
its thickness on timescales of $\sim 5$ Gyr. However, if the disc
indeed lies in the planes perpendicular to the long or short axis of a
triaxial halo, then it could survive and preserve its thinness over
cosmological timescales. 

\begin{figure*}
\begin{center}
\includegraphics[scale=0.2]{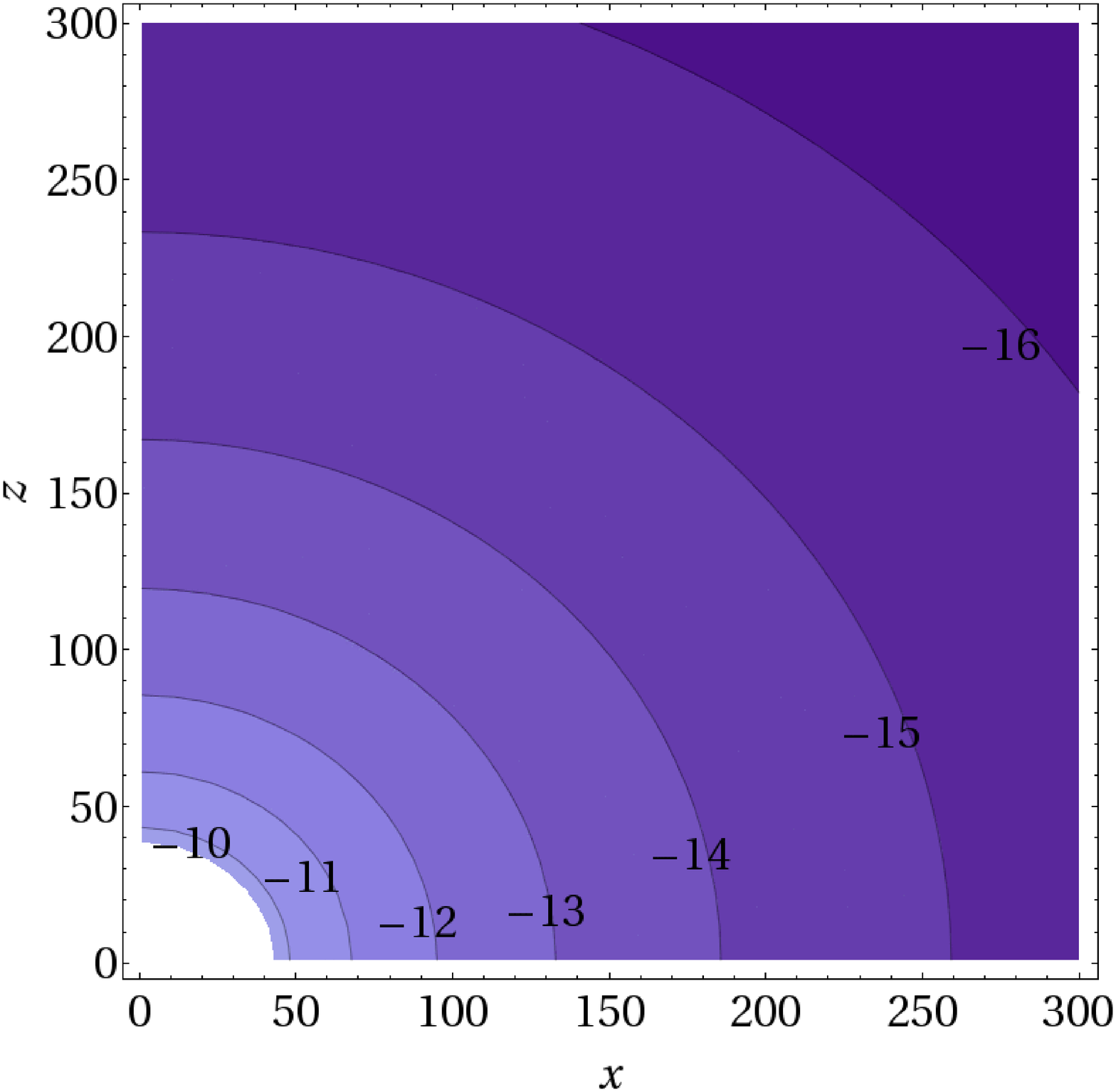}
\includegraphics[scale=0.2]{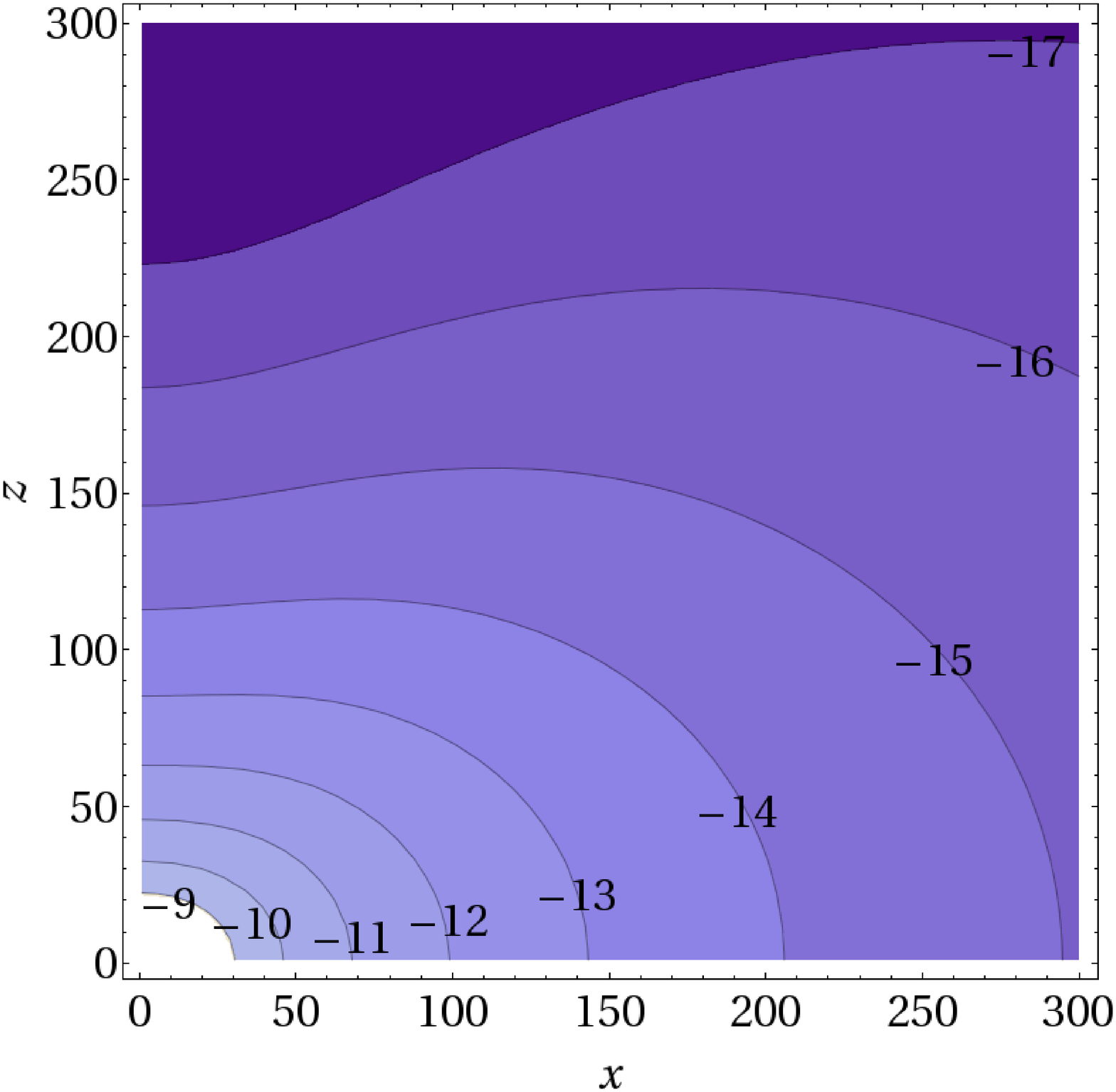}
\includegraphics[scale=0.2]{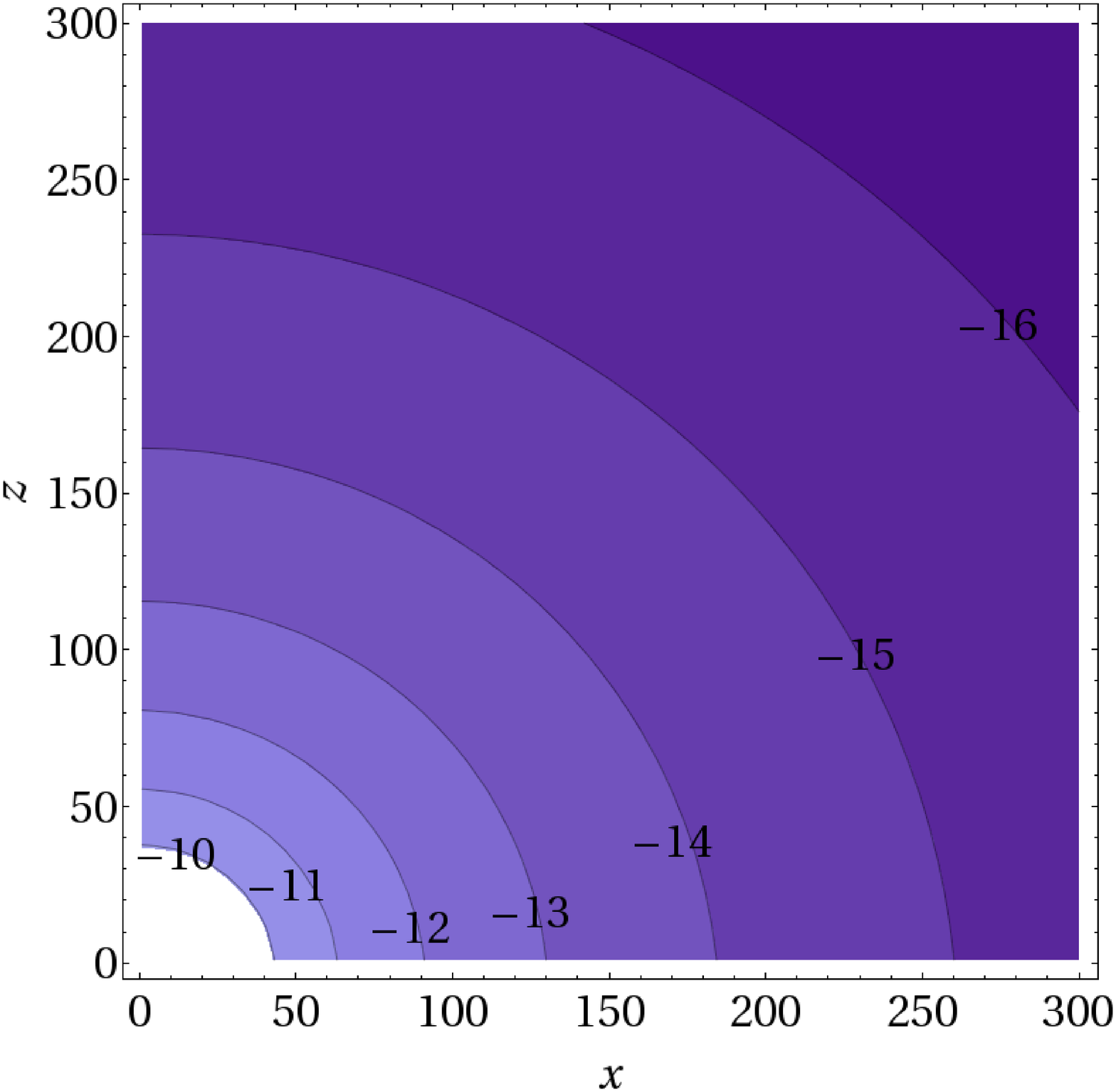}
\end{center}
\caption{Comparison of isodensity surfaces using three popular ways of
  modelling asphericity.  In the left panel, the density
  contours are elliptical, in the middle panel the equipotentials are
  elliptical and in the right panel the flattening is produced by
  spherical harmonics. (The contour labels are in logarithm of the 
  density with units of $10^{10} M_\odot$ kpc$^{-3}$).}
\label{fig:comps}
\end{figure*}
\begin{figure*}
\begin{center}
\includegraphics[scale=0.2]{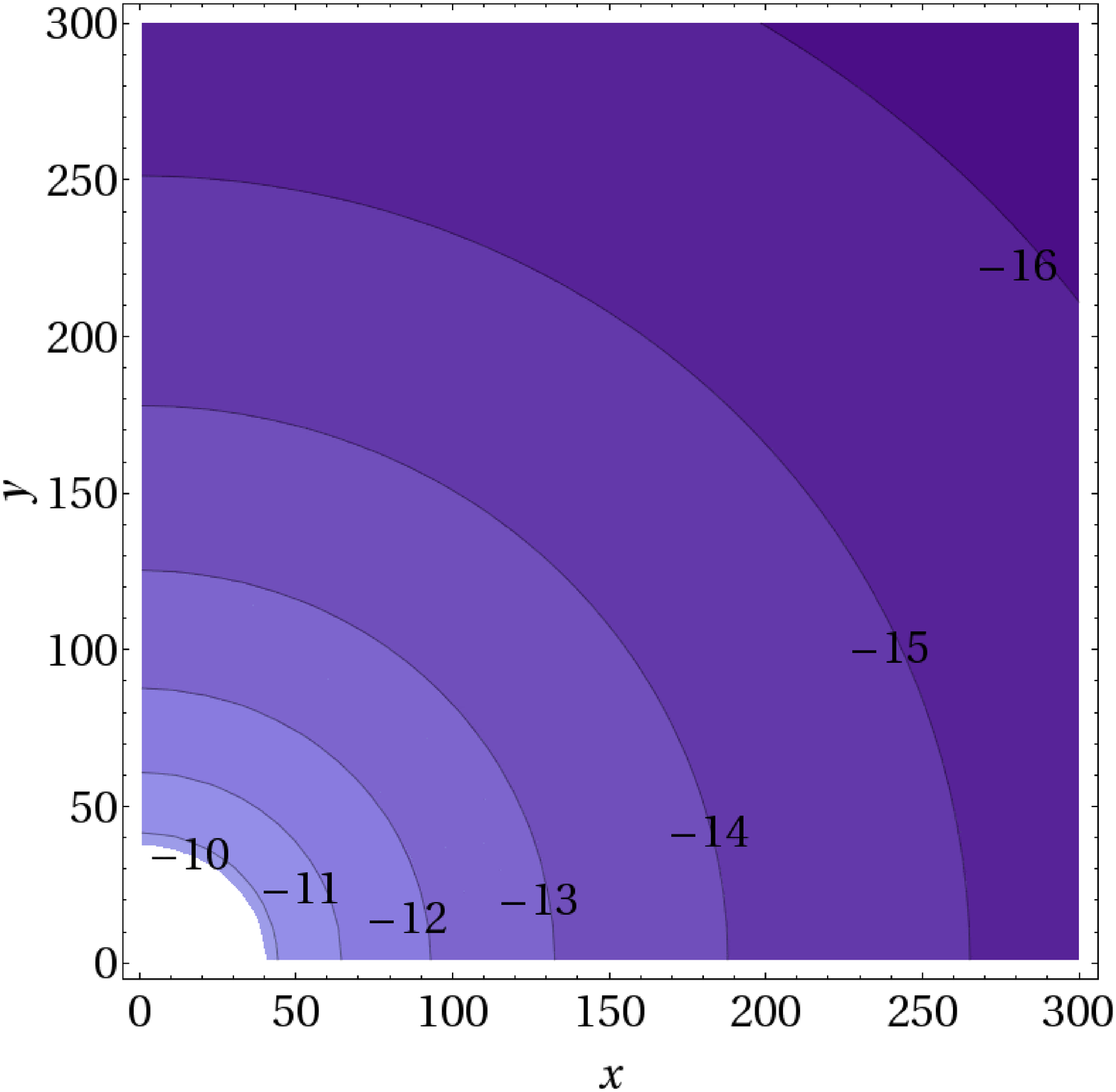}
\includegraphics[scale=0.2]{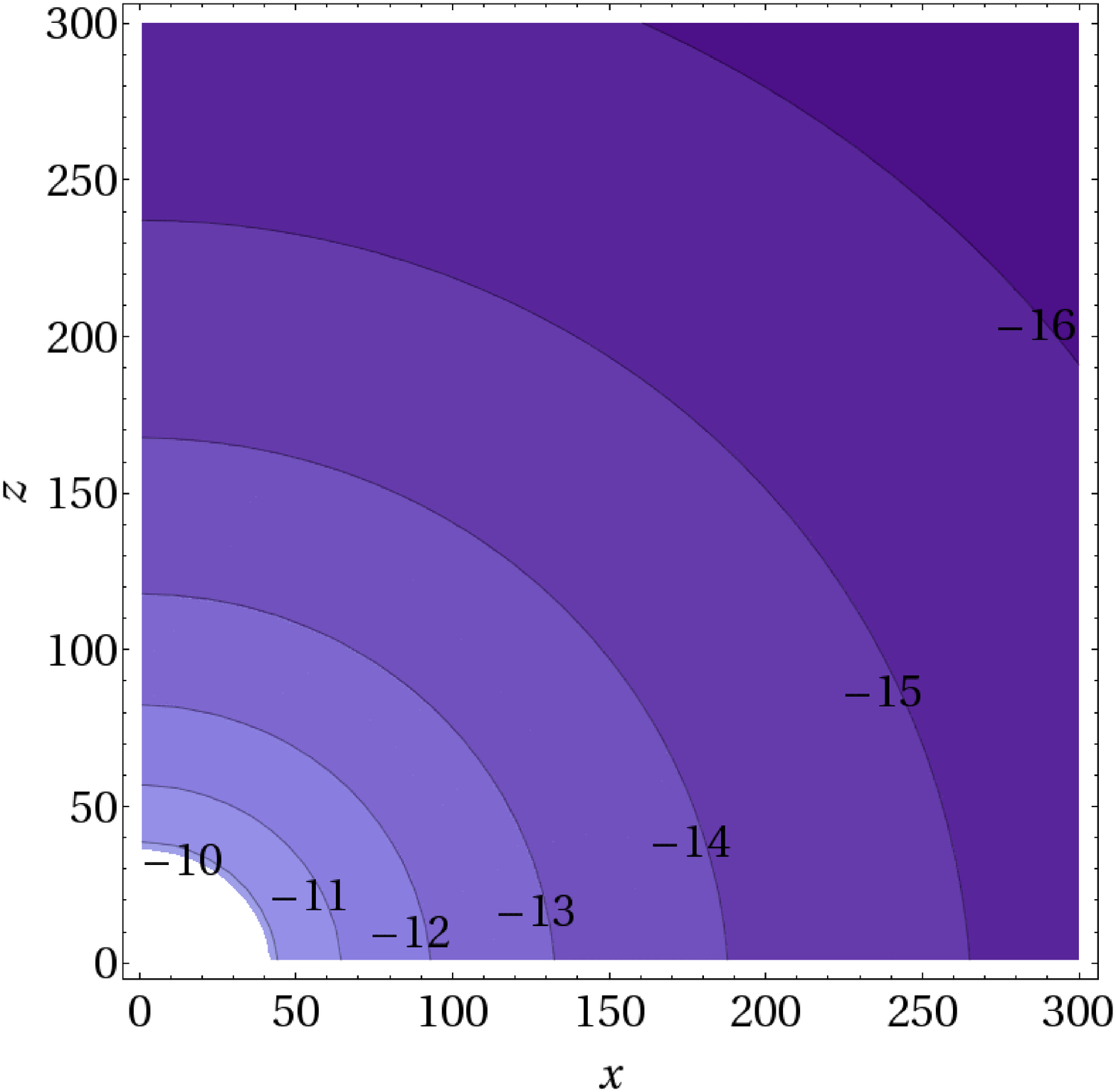}
\includegraphics[scale=0.2]{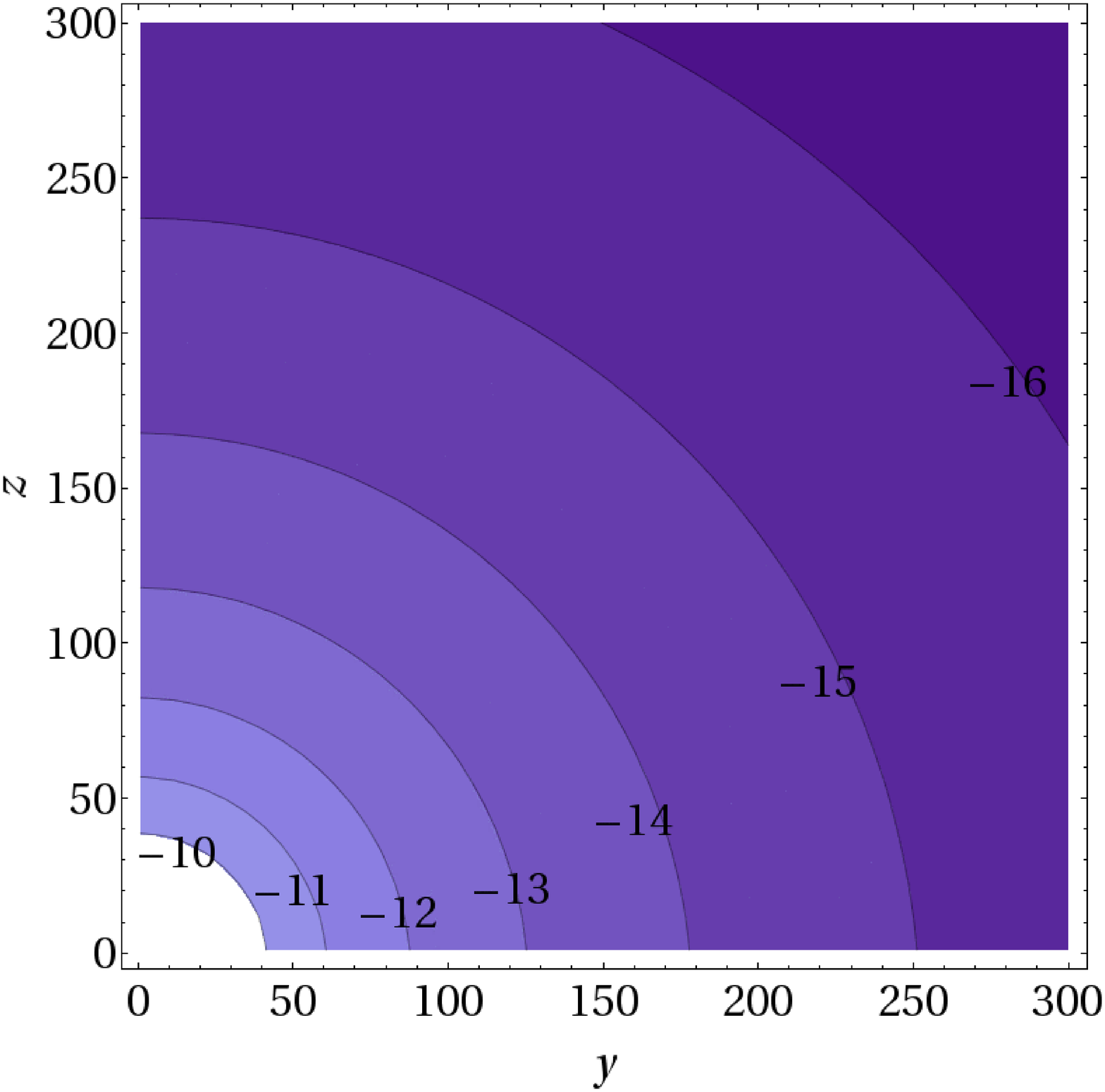}
\end{center}
\caption{Contour plots of density in the principal planes of a
  triaxial Navarro-Frenk-White model. The $x$ axis is the longest, the
  $y$ axis the intermediate and the $z$ axis the shortest of the
  triaxial figure. The central value of the axis ratios are $p_0 =
  (b/a)_0 = 0.9$ and $q_0 = (c/a)_0 =0.8$. At large radii, the model
  becomes rounder with axis ratios $p_\infty = (b/a)_\infty = 0.95$
  and $(c/a)_\infty = q_\infty= 0.9$.}
\label{fig:contours}
\end{figure*}
\begin{table*}
\begin{tabular}{l c c c c c c c c c c r}
\hline
Model & Description & M31 mass & c & {$q_0$} & {$q_\infty$} &
$p_0$ & $ p_\infty$ & {$\psi$} & Milky Way & $\th$ & rms scatter  \\
Number & & (in $10^{12} M_\odot$) & & & & & & (in deg) & Included ? & (in Gyr) & (in kpc) \\
\hline
1& Prolate & 1.0 & 10.0 & 1.2 & 1.1 & 1.0 & 1.0 & 0.0 & Yes & Stable & 3.6  \\
2& Oblate & 1.0 & 10.0 & 0.8 & 0.9 & 1.0 & 1.0 & 90.0 & Yes & Stable & 4.8 \\
3& Oblate & 1.0 & 10.0 & 0.8 & 0.9 & 1.0 & 1.0 & 77.5 & Yes & 7.02 & 9.8 \\
4& Oblate & 1.0 & 10.0 & 0.8 & 0.9 & 1.0 & 1.0 & 45.0 & No & 5.47 & 19.4 \\
5& Triaxial & 1.0 & 10.0 & 0.8 & 0.9 & 0.9 & 0.95 & 90.0 & No & Stable & 3.6 \\
6& Triaxial & 1.0 & 10.0 & 0.8 & 0.9 & 0.9 & 0.95 & 77.5 & No & 6.99 & 9.4 \\
\hline
\end{tabular}
\caption{Properties of the models. The M31 halo has an NFW profile
  with mass $M$ and concentration $c$. The oblateness or prolateness
  is controlled by $q_0$ and $q_\infty$, whilst the triaxiality is
  controlled by $p_0$ and $p_\infty$.  The angle of inclination of the
  disc of satellites to the pole ($z=0$) is given by $\psi$. So, for
  example, in model 2, 90$^\circ$ means that the disc of satellites
  lies in the equatorial plane, while in model 3, $77^\circ$ means that the
  disc of satellites is coplanar with the stellar disc of M31. When
  the disc is unstable, then the e-folding time $\th$ is also given.
  The rms vertical scatter at the end of the simulation is recorded in
  kpc.}
\label{tab:models}
\end{table*}

\section{Dark Halo Models}

For some applications, oblate or prolate NFW haloes suffice. So we
deal with the simpler case in subsection 2.1, before
passing to the triaxial case in subsection 2.2.

\subsection{Oblate and Prolate Navarro-Frenk-White Halos}

To provide a simple and flexible family of flattened halo models, we
consider potentials of the form
\begin{equation}
\psi_{\rm axi}(r,\theta) = \psi_{\rm NFW}(r) - 4\pi G {\rho_1 r_1^3r \over (r+r_1)^2} Y^0_2(\theta),
\label{eq:pot}
\end{equation}
where ($r,\theta$) are spherical polar coordinates. $Y^0_2(\theta) =
\frac{3}{2}\cos^2 \theta - \frac{1}{2}$ is the familiar second
spherical harmonic, which shortens (or lengthens) the $z$ axis of the
density distribution, whilst lengthening (or shortening) the $x$ and
$y$ axes to give an oblate (or prolate) figure. The flattened density
is
\begin{equation}
\rho_{\rm axi}(r,\theta) = \rho_{\rm NFW}(r) - 2 \rho_1r_1^3 {3r^2 +
  8rr_1 + 2r_1^2 \over r(r+r_1)^4 } Y^0_2(\theta).
\label{eq:dens}
\end{equation}
Notice that the density still falls off like $r^{-1}$ at small radii
and $r^{-3}$ at large radii, though this model is now oblate or
prolate. There are two additional parameters $\rho_1$ and $r_1$, which
control the normalisation and lengthscale of the asphericity and hence
ellipticity of the density contours at large and small radii. We find
\begin{equation}
{\rho_0 \rs^3 \over \rho_1 r_1^3} = 3 \Biggl( {2 + q_\infty^3 \over
  1 - q_\infty^3} \Biggr),
\label{eq:ax0}
\end{equation}
\begin{equation}
{r_1^2 \over \rs^2} = {2 \over 3} \Biggl( {2 + q_0 \over
  2+ q_\infty^3} \Biggr) \Biggl( { 1- q_\infty^3 \over 1 -q_0 }
\Biggr),
\label{eq:ax1}
\end{equation}
where $q_0$ and $q_\infty$ are the axis ratios of the isodensity
contours near the center and at large radii.  These formulae provide a
simple way to fix the parameters of the model. First, $\rho_0$ and
$\rs$ are chosen to lie on the mass-concentration relationship derived
from cosmological simulations~\citep[see e.g.,][]{Mac07}. Then,
$\rho_1$ and $r_1$ are chosen using eqns(\ref{eq:ax0}) and
(\ref{eq:ax1}) to give the desired axis ratio near the centre $q_0$
and at large radii $q_\infty$.

Fig.~\ref{fig:comps} shows a comparison between the three popular ways
of flattening a spherical model. Either the density or potential
contours can be made elliptical (left and middle panels) or spherical
harmonics can be added to the potential (right). Note that making the
potential elliptical often results in unphysical densities, as
here. By contrast, adding spherical harmonics retains a simple
potential, but gives much more realistic contours.

\subsection{Triaxial Navarro-Frenk-White Halos}

With a little more effort, fully triaxial NFW haloes can be obtained
by adding a further spherical harmonic, $Y^2_2(\theta,\phi) = 3 \sin^2
\theta \cos 2\phi$.  The density and potential become
\begin{equation}
\psi_{\rm tri}(r,\theta,\phi) = \psi_{\rm axi}(r,\theta) + 4\pi G
    {\rho_2 r_2^3r \over (r+r_2)^2} Y^2_2(\theta,\phi),
\label{eq:tripot}
\end{equation}
\begin{equation}
\rho_{\rm tri}(r,\theta,\phi) = \rho_{\rm axi}(r,\theta) 
+ 2 \rho_2r_2^3 {3r^2 +
  8rr_2 + 2r_2^2 \over r(r+r_2)^4 } Y^2_2(\theta,\phi).
\label{eq:tridens}
\end{equation}
The figure is now truly triaxial with the $x$ axis as the longest and
the $z$ axis the shortest. Just as for the oblate and prolate models,
the parameters can be fixed by choosing the axis ratios 
at small radii $q_0, p_0$:
\begin{equation}
{\rho_0 \rs^3 \over \rho_1 r_1^3} = 6  {1 +p_\infty^3 + q_\infty^3
  \over  1 + p_\infty^3 - 2q_\infty^3},
\end{equation}
\begin{equation}
{\rho_0 \rs^3 \over \rho_2 r_2^3} = 12{1+p_\infty^3 + q_\infty^3
  \over  (1 - p_\infty^3)},
\end{equation}
and at large radii $q_\infty,p_\infty$:
\begin{equation}
{r_1^2 \over \rs^2} = {2 \over 3}  {(1+p_0+q_0)(1+p_\infty^3-2q_\infty^3)
  \over (1+ p_0 -2 q_0)(1+p_\infty^3 + q_\infty^3) },
\end{equation}
\begin{equation}
{r_2^2\over \rs^2} = {2\over 3}
{(1+p_0+q_0)(1-p_\infty^3)
\over  (1-p_0)(1+p_\infty^3+q_\infty^3)}.
\end{equation}
Fig.~\ref{fig:contours} shows isodensity contours in the principal planes
of a triaxial model with $p_0 = 0.9, p_\infty = 0.95$ and $q_0 = 0.8,
q_\infty= 0.9$. The contours are ellipse-like curves, which befits
models of dark halos and elliptical galaxies. If the axis ratios are
made smaller, then the contours can become dimpled. This can be useful
for some applications, but -- if not desired -- it can be avoided by
the addition of further harmonics.  The positivity of the density is
not guaranteed and needs to be checked {\it a posteriori}, but there
is a generous range of parameter space for which the potential-density
pair generate triaxial cosmologically inspired halos with everywhere
positive density.

\subsection{Closed Orbits}

In the absence of figure rotation, closed orbits exist in the planes
perpendicular to the long and the short axes of a triaxial
potential~\citep{He79}. These can naturally form the backbone of discs
-- whether of cold gas or satellite galaxies -- and so are of special
interest here. We might expect that the closed orbits in the planes
perpendicular to the long and short axes can support stable and
long-lived discs.

The properties of the closed orbits can be found by epicyclic theory.
For example, in the ($x,y$) plane, the potential has the form
\begin{equation}
\psi_{\rm tri}(r,{\pi/2},\phi) = V_0(r) + V_2(r) \cos 2\phi,
\label{eq:planar}
\end{equation}
with
\begin{equation}
V_0(r) = \psi_{\rm NFW}(r) +  2\pi G {\rho_1 r_1^3r \over
  (r+r_1)^2},\quad
V_2(r) = 12\pi G
    {\rho_2 r_2^3r \over (r+r_2)^2}.
\end{equation}
A similar result to eq~(\ref{eq:planar}) holds good in the ($y,z$)
plane.

The closed orbits are confined to the plane and oriented opposite to
the longer axis. To lowest order, they are ellipses given by
\begin{equation}
r = r_0\left( 1 - \frac{1}{2} \epsilon \cos 2\phi \right),
\end{equation}
where $\epsilon$ is the orbital ellipticity $1 - r_{\rm min}/r_{\rm
  max}$. By means of the equations in \citet{Ge86}, we find that the
ellipticity varies with radius like
\begin{equation}
\epsilon = {2\over r(4\Omega^2 - \kappa^2 )}\left( {2V_2\over r} + 
{d V_2\over dr} \right),
\end{equation}
where $\Omega$ and $\kappa$ are the circular and epicylic frequencies
respectively. The closed orbits are circular at large radii, but they
become increasingly elliptic at small radii. For example, in the ($x,y$)
plane, they reach a limiting ellipticity
\begin{equation}
\epsilon \rightarrow {36 r_2 \rho_2 \over r_s \rho_0 - r_1 \rho_1},
\end{equation}
as $r \rightarrow 0$.

We might therefore expect discs of satellites to be possible in
triaxial potentials only in the planes perpendicular to the long and
short axes. For all other orientations, we expect thickening of the
disc as the orbits respond to gravitational torques  

\begin{figure}
\begin{center}
\includegraphics[scale=0.3]{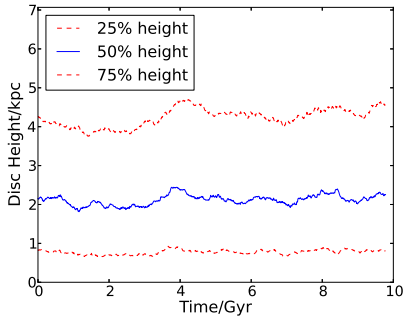}
\end{center}
\caption{The evolution of the quarter-mass, half-mass and
  three-quarter-mass heights of the disc in a spherical NFW profile,
  evolved for 10 Gyrs. The vertical height of the disc remains
  unchanged, validating our way of setting up initial conditions.}
\label{fig:confirm}
\end{figure}
\begin{figure}
\begin{center}
\subfigure[Model 3]{\includegraphics[scale=0.3]{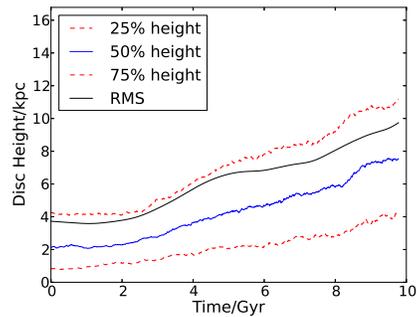}}
\subfigure[Model 4]{\includegraphics[scale=0.3]{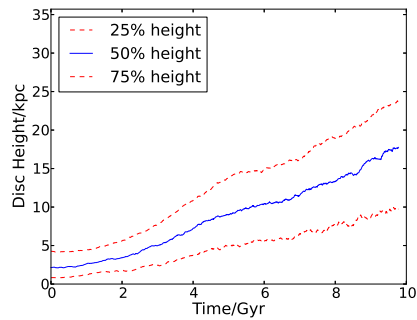}}
\end{center}
\caption{Scale height evolution of the disc of satellites in an oblate
  NFW profile tilted at 77.5$^\circ$ (upper panel). This means that
  the principal axes of the halo are aligned with the stellar disc of
  M31. The full black line shows the evolution of the rms scatter,
  which is the quantity measured by observers. It typically lies
  between the half-mass and three-quarter-mass scaleheights.  Also
  shown is a model in which the disc of satellites is tilted at
  45.0$^\circ$ to the halo (lower panel). In both cases, the thin disc
  becomes fat over timescales $\sim 6$ Gyr. This is a general result
  whenever the disc of satellites is misaligned with the principal
  planes.}
\label{fig:evolunstab}
\end{figure}

\section{The Disc of Satellites}

\subsection{Initial Conditions}

The satellite galaxies lie in the near-Keplerian outer parts of the
M31 halo. We first develop a way of setting up approximate initial
conditions for thin discs in the outer parts of a triaxial halo.

First, Keplerian semi-major axes $a$ and eccentricities $e$ are drawn
from uniform distributions of 100-200 kpc and 0.0-0.5 respectively,
with a random line of apsides $\thetaa$ in the disc ($x,y$) plane. 
This gives the Cartesian coordinates at apocentre as
\begin{equation} \Bigl( x_{\rm a}, y_{\rm a}\Bigr) = a (1+e)
  \Bigl( \cos\thetaa, \sin\thetaa \Bigr),
\end{equation}
A Keplerian ellipse has a tangential velocity at apocentre of
\begin{equation} v_{\phi} = \sqrt{\frac{G M (1-e)}{a(1+e)}}, 
\end{equation}
giving planar velocities of
\begin{equation} \Bigl( v_{x}, v_y \Bigr) = \frac{v_{\phi}}{a (1+e)}
  \Bigl(y_{\rm a}, -x_{\rm a }\Bigr).
\end{equation}
So far, the disc is razor-thin. To thicken up the disc, orbital
inclinations $i$ are chosen uniformly in the range 0-3$^\circ$.  The
vertical motion at apocentre in a near-Keplerian potential can then be
approximated to first order as
\begin{equation} \ddot{z} = \frac{-G M z}{r^3} = \omega^2_z z, \qquad
\omega_z = \sqrt{\frac{G M}{a^3 (1+e)^3}}.
\end{equation}
This allows us to choose a random phase, \emph{$\phi$} from 0 to
$2\pi$, such that
\begin{equation} z_{\rm a} = a(1+e) \sin i \cos{\phi}, \qquad v_{z} =
  a(1+e)\omega_z \sin i \sin{\phi}. 
\end{equation}
This completes the specification of the apocentres of our orbits. Of
course, we wish our satellites to begin at a random point along their
orbits. The Keplerian orbital period is $P = \sqrt{4 \pi^2 a^3/(G
  M)}$.  A random number is generated in the range 0 to $P$, and the
orbit evolved forward for this time from apocentre. The properties of
the satellite at this time provide the initial phase-space coordinates
at the start of our simulations

A disc created in this fashion is extended and thin.
Fig.~\ref{fig:confirm} shows the evolution of such a disc in a
spherical Navarro-Frenk-White potential. It starts out with an rms
vertical scatter of $3.7$ kpc, and the coloured lines show the
evolution of scaleheights enclosing a quarter, a half and
three-quarters of the satellites.  For a disc with a vertical Gaussian
profile, the half-mass thickness is comparable to the rms scatter.
After 10 Gyr, the disc has a vertical scatter of $4.5$ kpc, only
slightly increased from its starting value. It is much thinner than
that claimed for the disc of satellites around M31~\citep{Ib13}.  This
confirms the stability of the disc and validates our algorithm for
initial conditions.

\subsection{The Tidal Effects of the Milky Way Galaxy}

Finally, we include the effect of tidal forces from the Milky Way
galaxy on the disc of satellites.  The present day separation of the
Milky Way and M31 is $\sim 770$ kpc, with the Milky Way approximately
coplanar with the observed disc around M31. As this distance was still
greater in the recent past, the Milky Way galaxy does not need to be
modelled in detail in our simulations.  Accordingly, the Milky Way
halo is taken as a spherical NFW profile with a halo mass {$M$} of
{$10^{12} M_{\odot}$} and a concentration $c$ of $10$. The evolution
of the separation between the Milky Way and M31 can be modelled with
the Timing Argument~\citep{Pe93}, which assumes that the mutual
gravitation of the two galaxies caused them to decouple from the
Hubble flow at high redshift. Specifically, we use eqn (10) in
\citet{vdM08}, with our simulations running from t = 3.95 Gyr to the
present day at t = 13.73 Gyr.

\subsection{Results}

We performed orbit integrations for a population of 1000 satellite
galaxies modelled as test particles in a variety of oblate, prolate
and triaxial haloes listed in Table~\ref{tab:models}. 

To study the vertical structure, we track the evolution of the
quarter-mass, half-mass and three-quarter-mass heights over 10 Gyr.
If the disc thickens, then the growth of the disc half-mass height
with time is fitted to an exponential functional form, $z = z_0\exp
t/\th$, and the $e$-folding timescale $\th$ is reported in
Table~\ref{tab:models}.

The first two models record simulation results for oblate and prolate
E2 haloes. The disc of satellites lies either in the polar (model
1) or equatorial planes (model 2).  The tidal effects of the Milky Way are
incorporated, as well as of course the torques from the asphericity of
the haloes. However, these forces do not disturb the serenity of the
discs, which are long-lived. This result is straightforward to
understand. Circular orbits in the equatorial plane of either oblate
or prolate dark haloes are stable, and can sire families of orbits
which are long-lived and populate a disc of satellites. Equally, the
non-circular, but closed and periodic, orbits in the meridional planes
of either oblate or prolate haloes can be used to build stable discs.

However, any misalignment of the principal planes of the oblate or
prolate halo with the disc of satellites causes the disc to
fatten. This is the case in models 3 and 4 in Table~\ref{tab:models}.
The upper panel of Fig.~\ref{fig:evolunstab} shows an E2 oblate halo,
in which the equatorial plane is now aligned with the stellar disc of
the M31 galaxy, which is at an inclination of $77.5^\circ$\citep[see
  e.g.,][]{Ho92}. The disc of satellites no longer lies in the
equatorial plane. This is a natural model to use if we believe that
the shape of the dark matter halo does not twist or misalign with
increasing radius. Notice that the disc of satellites is now no longer
a stable structure -- the disc height doubles within $\sim 7$ Gyr. In
a flattened halo, the only planar orbits are confined to the principal
planes. Once the halo is misaligned with the disc of satellites, then
torques from the quadrupole and higher moments cause the disc to
thicken. The bottom panel of Fig.~\ref{fig:evolunstab} again shows an
E2 oblate halo, but the misalignment is now $45^\circ$. This dark halo
is misaligned both with the stellar disc and the disc of
satellites. The disc is fattened more quickly, with a doubling of the
disc height in $\sim 5.5$ Gyr. At the end of the simulations, the rms
scatter is $\sim 19$ kpc, comparable to the rms scatter claimed for
the disc of satellites.

Models 5 and 6 of Table~\ref{tab:models} show the effects of
incorporating triaxiality. In model 5, the halo is aligned so that the
$x$ axis points to the Milky Way and the disc of satellites occupies
the ($x,y$) plane. As noted in our discussion of closed orbits, stable
closed orbits can exist in the planes perpendicular to the long and
the short axes of a triaxial halo. The simulation demonstrates that
the disc can survive unthickened for timescales $\sim 10$ Gyr. Once
however, the halo is misaligned with the disc of satellites, thickening
does occur. For example, even the modest misalignment of $12.5^\circ$
of model 6 gives a doubling of the disc height on a timescale of $\sim
7$ Gyr.

\section{Conclusions}

We have presented a simple triaxial generalization of the
Navarro-Frenk-White model. It has an elementary potential-density
pair. The parameters of the model can be chosen to lie on the
cosmological mass-concentration relation, and tuned to give any
desired axis ratios at large and small radii. So, it is easy to build
cosmologically-inspired haloes with radially varying triaxiality,
tailored for any purpose -- for example, oblate in the centre but
prolate in the outer parts, or triaxial in the centre and round in the
outer parts.

We have used the haloes to examine the longevity of a vast, thin disc
of satellites. This is motivated by the observational claims
of~\citet{Ib13}, who have surveyed the outer parts of M31. Such a disc
is in general not long-lived, and thickens due to torques and tidal
forces.  The disc of satellites therefore provides a powerful
constraint on the shape and orientation of the M31 dark halo.  We find
that there are the following three possibilities for a long-lived
disc:

\medskip
\noindent
[1] If the M31 dark halo is spherical, then a thin disc of satellites
can survive for timescales of the order of 10 Gyr.

\medskip
\noindent
[2] If the M31 dark halo is oblate or prolate, and the disc of
satellites lies in the equatorial or polar plane of the spheroidal
potential, then it is also thin and long-lived. The same holds true
for triaxial haloes, provided the disc lies in the plane perpendicular
to the long or short axis. For M31, such a configuration would require
that the principal axes of the dark halo at large radii are misaligned
with the axes of the stellar disc of the galaxy itself. This requires
the M31 potential to twist on moving outwards from the
baryon-dominated central parts to the outer halo. Such misalignment is
often observed in numerical simulations~\citep[see e.g.,][]{De11}.

\medskip
\noindent
[3] For all other orientations, the disc of satellites must thicken
and its vertical height grows exponentially with an e-folding time of
$\sim 5$-$7$ Gyr. If its present rms scatter is $\sim 12$ kpc, then it
must have been formed with a scaleheight that is astonishingly thin,
perhaps only $2$ or $3$ kpc despite being hundreds of kpc across. This
seems implausible. Nonetheless, the problem could be ameliorated if
there are more satellites associated with the disc of satellites than
suggested by \citep{Ib13}, and so its present-day thickness could be
greater.

\section*{Acknowledgements}
AB thanks the Science and Technology Facilities Council (STFC) for the
award of a studentship. VB is supported by the Royal Society. The
referee is thanked for a helpful and constructive report.


\begin{thebibliography}{}

\bibitem[Allgood et al.(2006)]{Al06} Allgood, B., Flores, R.~A.,
  Primack, J.~R., et al. 2006, MNRAS, 367, 1781

\bibitem[Conn et al.(2012)]{Co12} Conn, A.R., et al., 2012, ApJ, 758,
  11

\bibitem[Conn et al.(2013)]{Co13} Conn, A.R., et al., 2013, ApJ, 766
  120 

\bibitem[Deason et al.(2011)]{De11} Deason, A.J., McCarthy, I.G.,
  Font, A.S., et al.\ 2011, MNRAS, 415, 2607

\bibitem[de Zeeuw \& Carollo(1996)]{Z96} de Zeeuw, P.T., Carollo,
  C.M., 1996, MNRAS, 281, 1333

\bibitem[Dubinski(1994)]{Du94} Dubinski, J. 1994, ApJ, 431, 617

\bibitem[Fusi Pecci et al.(1995)]{Fu95} Fusi Pecci, F., Bellazinni,
  M., Cacciair, C., Ferraro, F.R., 1995, AJ, 110, 1664

\bibitem[Gerhard \& Vietri(1986)]{Ge86} Gerhard, O.E., Vietri, M., 1986,
  MNRAS, 223, 377

\bibitem[Heiligman \& Schwarzschild(1979)]{He79} Heiligman, G., 
  Schwarzschild, M. 1979, ApJ, 233, 872

\bibitem[Hernquist \& Quinn(1989)]{He89} Hernquist, L., Quinn, P.J.,
  1989, ApJ, 342, 1

\bibitem[Hodge(1992)]{Ho92} Hodge, P.W. 1992 The Andromeda Galaxy,
  Kluwer, Dordrecht

\bibitem[Ibata et al.(2013)]{Ib13} Ibata, R.A., Lewis, G.F., Conn,
  A.R., et al.\ 2013, Nat, 493, 62

\bibitem[Jing \& Suto(2002)]{Ji02} Jing, Y.~P., Suto, Y.\ 2002,
  ApJ, 574, 538

\bibitem[Koch \& Grebel(2006)]{Ko06} Koch, A., Grebel E., 2006, AJ,
  1405
 
\bibitem[Kazantzidis et al.(2004)]{Ka04} Kazantzidis, S., Kravtsov,
  A.~V., Zentner, A.~R., et al.\ 2004, ApJ, 611, L73


\bibitem[Maccio et al.(2007)]{Mac07} Maccio, A.V., Dutton A.A., van
  den Bosch, F., Moore, B., Potter, D., Stadel J., 2007, MNRAS, 378,
  55




\bibitem[Mo, van den Bosch \& White(2010)]{Mo10} Mo, H.J., van den
  Bosch, F., White, S., 2010, Galaxy Formation and Evolution,
  Cambridge University press, Cambridge, chap. 7

\bibitem[Peebles(1993)]{Pe93} Peebles, P.J.E., 1993, Principles of
    Physical Cosmology, Princeton University Press, Princeton


\bibitem[Schwarzschild(1979)]{Sch79} Schwarzschild, M. 1979, ApJ, 232,
  236

\bibitem[Schwarzschild(1993)]{Sch94} Schwarzschild, M. 1993, ApJ, 409,
  563

\bibitem[van der Marel \& Guhathakurta(2008)]{vdM08} van der Marel,
  R.~P., Guhathakurta, P.\ 2008, ApJ, 678, 187



\bibitem[Zemp et al.(2012)]{Ze12} Zemp, M., Gnedin, O.~Y., Gnedin,
  N.~Y., \& Kravtsov, A.~V.\ 2012, ApJ, 748, 54

\end{thebibliography}
\end{document}